\documentclass[preprint]{revtex4}  
\usepackage{amsmath}
\usepackage{amssymb}
\usepackage{amsbsy}
\usepackage{graphicx}
\usepackage{wasysym}
\usepackage[dvips]{epsfig}
\usepackage{psfrag}
\usepackage[ansinew]{inputenc}
\usepackage[T1]{fontenc}
\usepackage{lmodern}
\usepackage{xcolor}

\newcommand{\ds}{\displaystyle}
\newcommand{\ul}{\underline}

\newcommand{\p}{\partial}

\newcommand{\epsi}{\varepsilon}

\newcommand{\bra}[1]{\left\langle{#1}\right|}

\newcommand{\braket}[2]{\left\langle{#1} \,|\, {#2} \right\rangle}

\newcommand{\op}[1]{\mathsf #1}
\newcommand{\set}[1]{\mathcal{#1}}

\renewcommand{\vec}[1]{\mbox{\boldmath $ #1 $}}
\newcommand{\R}{\mathbb{R}}
\newcommand{\N}{\mathbb{N}}

\def\gu{\;{\lower0.3ex\hbox{$\buildrel > \over {\scriptstyle \sim}$}}\;}
\def\lu{\;{\lower0.3ex\hbox{$\buildrel < \over {\scriptstyle \sim}$}}\;}

\def\XXint#1#2#3{{\setbox0=\hbox{$#1{#2#3}{\int}$}
     \vcenter{\hbox{$#2#3$}}\kern-.5\wd0}}

\begin{document}

\title{The spherical multipole resonance probe:\\ kinetic damping in its spectrum}
\author{J.~Oberrath}
\affiliation{Leuphana University L\"uneburg,\\Institute of Product and Process Innovation,\\ Universit\"ätsallee 1, 21335 L\"uneburg, Germany}
\date{\today}
\begin{abstract}

The multipole resonance probe is one of the recently developed measurement devices to measure plasma parameter like electron density and temperature based on the concept of active plasma resonance spectroscopy. The dynamical interaction between the probe and the plasma in electrostatic, kinetic description can be modeled in an abstract notation based on functional analytic methods. These methods provide the opportunity to derive a general solution, which is given as the response function of the probe-plasma system. It is defined by the matrix elements of the resolvent of an appropriate dynamical operator. Based on the general solution a residual damping for vanishing pressure can be predicted and can only be explained by kinetic effects. Within this manuscript an explicit response function of the multipole resonance probe is derived. Therefore, the resolvent is determined by its algebraic representation based on an expansion in orthogonal basis functions. This allows to compute an approximated response function and its corresponding spectra, which show additional damping due to kinetic effects.

\end{abstract}
\maketitle

\section{Introduction}
Active plasma resonance spectroscopy (APRS) denotes a concept of plasma diagnostics in which a radio frequent signal is coupled into a plasma via an electrical probe and the frequency dependent response is recorded. Based on a mathematical model plasma parameter like electron density and / or electron temperature can be determined from detected resonances. 

This concept was first investigated with the so called \textit{resonance probe} by Takayama, Ikegami, and Miyazaki \cite{takayama1960} in the year 1960. Since this year several different designs were invented, analyzed, and characterized \cite{levitskii1963, harp1964, fejer1964, crawford1964, dote1965, lepechinsky1966, buckley1966, balmain1966, davis1966, messiaen1966, mckeown1967, waletzko1967, kostelnicek1968, hellberg1968, li1970, cohen1971, tarstrup1972, aso1973, aso1973b, bantin1974, meyer1975, vernet1975, stenzel1976, nakatani1976, kist1977, morin1991, booth2005, scharwitz2009, xu2009, xu2010, li2010, liang2011, schulz2015}. New interest about this concept gained during the last two decades, especially in the context of industrial compatible plasma diagnostics. For this purpose a specific class of APRS is sufficient: probes, which excite resonances below the electron plasma frequency $\omega_{\rm p}$. Thus, the dynamical interaction between the probe and the plasma can be modeled in electrostatic approximation and can be applied without calibration. Representatives of this class are, e.g., the impedance probe (IP) \cite{blackwell2005, hopkins2014}, the plasma absorption probe (PAP) \cite{sugai1999}, and the multipole resonance probe (MRP) \cite{lapke2008, lapke2011}, just to name a few. For the latter could be shown to satisfy the optimized design \cite{lapke2013}.  

These probes are intensively investigated, experimentally and theoretically. The theoretical works cover analytic and numerical methods and plasma models of different complexity. They have in common, that they focus just on a particular probe design. However, it is also of interest to study generic features of such probes independently of any particular realization. Such an analysis, based on Hilbert space methods, is presented in \cite{lapke2013}. In the description of the cold plasma model the main result was that, for any possible probe design, the spectral response function could be expressed as a matrix element of the resolvent of the dynamical operator. 

The analysis of the study of \cite{lapke2013} can be generalized in a full kinetic description and yields an abstract kinetic model of electrostatic resonances valid for all pressures \cite{oberrath2014}. The main result of this model is similar to the result of the fluid model, which is: for any possible probe design, the spectral response of the probe-plasma system can be expressed as a matrix element of the resolvent of the dynamical operator. In addition, it was generally shown that resonances in the spectra of APRS probes exhibit a residual damping in the limit of vanishing pressure which can only be explained by kinetic effects and not by Ohmic dissipation. 

\pagebreak  

To study generic features of APRS probes Hilbert space methods are the optimal choice, because they allow an analysis of the probe-plasma system without any specification of the geometry. These methods can also be applied to determine explicit spectra for a specific probe design. This was done in the fluiddynamical description for certain geometries of the IP, the MRP, and the planar MRP, respectively \cite{oberrath2014b, friedrichs2018}, to determine analytic solutions of the response function. It is impossible to derive an analytic solution of the response function in the kinetic description, but an efficient algorithm, as shown in \cite{oberrath2016, oberrath2018} for the parallel electrode probe (PEP) and the IP, can be applied to determine an adequate approximation. The spectra of these probes computed with this algorithm show kinetic damping as predicted by the general analysis.  

However, kinetically calculated spectra of the MRP are not presented, yet. Within this manuscript the author follows the approximation algorithm presented in \cite{oberrath2016} to compute kinetic spectra of the MRP. These spectra will be compared to that of the IP \cite{oberrath2018}. It will be shown, that the kinetic determined resonances of these probe show a similar relation to each other, like the resonances determined with a fluid model, but of course with additional kinetic damping. 


\section{Model of the idealized multipole resonance probe}\label{sec:model}  
\begin{figure}[h!]
\includegraphics[width=0.65\columnwidth]{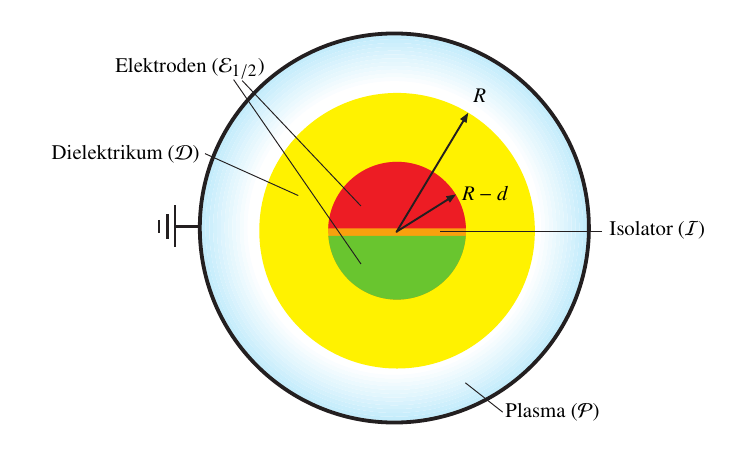} 
\vspace{-5mm}
\caption{Illustration of the idealized multiple resonance probe with powered electrodes ${\set E_{1/2}}$, dielectric $\set D$, and perturbed plasma $\set P$. The radius of the electrodes is $R-d$ and the thickness of the dielectric is $d$.}
\label{idealMRPkin}
\end{figure}
In fig.~\ref{idealMRPkin} the idealized MRP is depicted. It consists of two half-spherical electrodes $\set E_{1/2}$ of radius $R-d$, which are surrounded by a dielectric $\set D$ of thickness $d$. Applying RF voltages $U_{1/2}$ at the electrodes, which are phase shifted by 180$^\circ$, the plasma will be dynamically disturbed in the surrounding of the probe. This perturbed plasma $\set P$ united with the dielectric is defined as the influence region of the probe $\set V=\set P\cup\set D$. In the general kinetic model 
of \cite{oberrath2014} an interface $\set F$ between the perturbed and unperturbed plasma is defined, which is treated in this manuscript as a grounded spherical surface at a large distance $R_{\infty}$ on the length scale of the Debye length $\lambda_{\rm D}$ (Theoretically it is located in an infinite distance.). 

To describe the dynamics of the probe-plasma system in $\set P$ the linearized and normalized Boltzmann equation in electrostatic approximation has to be applied. The geometry of the idealized MRP is symmetric with respect to a rotation axes in $z$ direction, if the insulation of the electrodes is assumed to be in $x,y$ plane. Due to that, the 6 dimensional distribution function reduces to 5 dimensions and depends on the radial distance $r\in[R,R_{\infty}]$, the polar angle $\vartheta\in[0,\pi]$, the absolute value of the velocity $v\in[0,\infty)$, the projection angle $\chi\in[0,\pi]$ of $\vec v$ to the $r$ direction, and the rotation angle $\psi\in[0,2\pi]$ of $\vec v$ around $r$. The Boltzmann equation in these variables is given by
\begin{align} 
\frac{\p g}{\p t}
&+ v\left(\cos(\chi)\frac{\p}{\p r}+\frac{\sin(\chi)\cos(\psi)}{r}\frac{\p}{\p\vartheta}\right)\left(g-\op\Phi-\sum_{n=1}^2 U_n\psi_n\right) 
 + \frac{\p \bar\Phi}{\p r}
   \left(\cos(\chi)\frac{\p g}{\p v}
   -\frac{\sin(\chi)}{v}\frac{\p g}{\p \chi}\right)
  \nonumber\\
&-\sin(\chi)\frac{v}{r}\left(\frac{\p g}{\p\chi}
 +\cot(\vartheta)\sin(\psi)\frac{\p g}{\p\psi}\right) 
 =\frac{\nu_0}{4\pi}\int_0^{2\pi}\int_0^{\pi} g \sin(\chi)\, d\chi\, d\psi -\nu_0  g \ .
\label{LinBE} 
\end{align}

$g$ represents the perturbed distribution function of the electrons and is defined in $\set P$ with homogeneous boundary conditions at the surface of the probe $g(R,\vartheta,v,\chi ,\psi , t)=0$ and the outer grounded surface $g(R_{\infty},\vartheta,v,\chi ,\psi ,t)=0$. Only pure elastic collisions between electrons and the neutral background are taken into account and the collision frequency $\nu_0$ is assumed to be constant.
 
Due to the electrostatic approximation the force term is given by the electrical potential, which is separated in three parts. $\op\Phi$ is called inner potential and is a linear functional of $g$. It obeys the  Poisson equation
\begin{equation}
 \frac{1}{r^2}\frac{\p}{\p r}\left(r^2\varepsilon(r)\frac{\p\op\Phi}{\p r}\right)
+\frac{\varepsilon(r)}{r^2\sin(\vartheta)}\frac{\p}{\p\vartheta}
 \left(\sin(\vartheta)\frac{\p\op\Phi}{\p\vartheta}\right)=
\left\{\begin{matrix}
0 & , & r\in\set D\\[1ex]
\ds \int_{\R^3} w\, g \,d^3v & , & r\in\set P
\end{matrix}\right.
\label{PoissonMRP}
\end{equation}
with homogeneous boundary conditions $\op\Phi(R-d,\vartheta,t)=\op\Phi(R_{\infty},\vartheta,t)=0$. The dielectric constant $\varepsilon(r)$ is defined as $\varepsilon(r)=\varepsilon_D$ in $\set D$ and $\varepsilon(r)=1$ in $\set P$. $w$ is a positive weighting function, which is defined as the negative derivative of the equilibrium distribution $F(\epsilon)$ with respect to the total energy $\epsilon=\frac{1}{2}v^2-\bar\Phi$ in equilibrium. If the equilibrium distribution is assumed to be Maxwellian, $w$ equals to
\begin{equation}
w(r,v)=\frac{1}{(2\pi)^{\frac{3}{2}}}e^{-\frac{v^2}{2}+\bar\Phi(r)}
\end{equation} 
where $\bar\Phi$ is 
the equilibrium potential as second part of the potential.


The third part of the potential is represented by the radio frequent excitation of the probe in terms of the electrode functions $\psi_{1/2}(r)$. They follow Laplace's equation
\begin{equation}
 \frac{\p}{\p r}\left(r^2\frac{\p}{\p r}\psi_{1/2}^{(\set{D}/\set{P})}\right)
+\frac{1}{\sin(\vartheta)}\frac{\p}{\p\vartheta}
 \left(\sin(\vartheta)\frac{\p}{\p\vartheta}\psi_{1/2}^{(\set{D}/\set{P})}\right)=0
\label{Laplace}
\end{equation}
and fulfill the boundary conditions $\psi_1^{(\set D)}(R-d,\vartheta)=1$ if $\vartheta\in[0,\frac{\pi}{2}]$, $\psi_1^{(\set D)}(R-d,\vartheta)=0$ if $\vartheta\in(\frac{\pi}{2},\pi]$, $\psi_2^{(\set D)}(R-d,\vartheta)=1$ if $\vartheta\in(\frac{\pi}{2},\pi]$, $\psi_2^{(\set D)}(R-d,\vartheta)=0$ if $\vartheta\in[0,\frac{\pi}{2}]$, and $\psi_{1/2}^{(\set P)}(R_{\infty},\vartheta)=0$ and the transition conditions $\psi_{1/2}^{(\set{P})}(R,\vartheta)=\psi_{1/2}^{(\set{D})}(R,\vartheta)$ and $\frac{\p}{\p r}\psi_{1/2}^{(\set{P})}(R,\vartheta)=\varepsilon_D\frac{\p}{\p r}\psi_{1/2}^{(\set{D})}(R,\vartheta)$.

\section{Inner admittance in terms of functional analysis}

Within the last section the model of the idealized MRP is defined. Applying the results of the general analysis presented in \cite{oberrath2014}, the inner current $i_1$ at electrode $\set E_1$ can be written in a Hilbert space notation as 
\begin{equation}
i_1=\sum_{n'=1}^2\bra{e_1}\left(i\omega-\op T_V-\op T_S\right)^{-1} e_{n'}\,\rangle\, U_{n'}
   =\sum_{n'=1}^2\ Y_{1n'}\,U_{n'}.
\label{current}
\end{equation}
The inner admittance $Y_{1n'}$ is determined by the scalar product of the excitation and observation vectors
\begin{equation}
e_{1/2}=v\left(\cos(\chi)\frac{\p}{\p r}+\frac{\sin(\chi)\cos(\psi)}{r}\frac{\p}{\p\vartheta}\right)\psi_{1/2} 
\end{equation} 
and the resolvent of the dynamical operator $\op{T}_V+\op{T}_S$
\begin{equation}
Y_{1n'}=\langle e_1 |(i\omega-\op{T}_V-\op{T}_S)^{-1} e_{n'}\rangle \ .
\label{responseMRP}
\end{equation}
$\op T_V$ and $\op T_S$ are the Vlasov and the collision operator, respectively. They, applied to a dynamical state vector $g$, are defined as follows:
\begin{eqnarray}
\op T_V g 
& = & v\left(\cos(\chi)\frac{\p}{\p r}
      +\frac{\sin(\chi)\cos(\psi)}{r}\frac{\p}{\p\vartheta}\right)\left(\op\Phi - g\right) 
      + \frac{\p \bar\Phi}{\p r}
        \left(\frac{\sin(\chi)}{v}\frac{\p g}{\p \chi}
             -\cos(\chi)\frac{\p g}{\p v}\right)\nonumber\\
&   & +\sin(\chi)\frac{v}{r}\left(\frac{\p g}{\p\chi}
      +\cot(\vartheta)\sin(\psi)\frac{\p g}{\p\psi}\right) 
 \label{VlasovOp}\ ,\\[1ex]
\op T_S g 
& = & \frac{\nu_0}{4\pi}\int_0^{2\pi}\int_0^{\pi} g \sin(\chi)\, d\chi\, d\psi -\nu_0  g   \ .  
\end{eqnarray}
The scalar product, which is applied in \eqref{responseMRP}, is of particular importance for Hilbert space methods, but a physical interpretation of the final results is just available if the scalar product is connected to the system dynamics. This connection is fulfilled by a scalar product motivated by the kinetic free energy $\mathfrak{F}$, which is defined for two dynamical state vectors $g'$ and $g$ as follows
\begin{align}
\langle g^\prime | g \rangle 
&=\braket{g'}{g}_{\set P}+\braket{g'}{g}_{\set V}\label{ScalarProduct}\\[1ex] 
&=\frac{1}{\sqrt{2\pi}}\int_0^\pi\int_{R}^{R_\infty}\int_0^\pi\int_0^\pi\int_0^\infty  
  g^{\prime*} g\, n\,e^{-\frac{v^2}{2}} \sin(\chi)\sin(\vartheta) v^2 r^2
  \, dv\, d\chi\, d\psi\, dr\, d\vartheta\nonumber\\[1ex]
&\quad +2\pi\int_0^\pi\int_{R-d}^{R_\infty}\varepsilon
 \left(r^2\frac{\p \mathsf\Phi^{\prime*}}{\p r}\frac{\p \mathsf\Phi}{\p r} 
 + \frac{\p \mathsf\Phi^{\prime*}}{\p\vartheta}\frac{\p \mathsf\Phi}{\p\vartheta}\right) \sin(\vartheta)\,dr\, d\vartheta\nonumber\ . 
\label{ScalarProduct}
\end{align}
By means of a complete orthonormal basis $\{a\}$ in the Hilbert space $Y_{1n'}$ can be expanded based on the scalar product \eqref{ScalarProduct}. The corresponding completeness relation can be entered twice into equation \eqref{responseMRP} and yields
\begin{equation}
Y_{1n'}
=\sum_{a'}\braket{e_1}{a'}\sum_{a}\langle a' |
\left(i\omega-\op{T}_V-\op{T}_S\right)^{-1} a\rangle\braket{a}{e_{n'}}\ .
\label{responseExp}
\end{equation}
Such an expansion with an orthonormal and complete basis equates to a vector-matrix-vector multiplication in which the resolvent is represented by its algebraic equivalent. To determine the algebraic representation of the resolvent, the inverse of the algebraic representation of $i\omega-\op T_V-\op T_S$ has to be calculated \cite{oberrath2014b}.


\section{Explicit expansion of the inner admittance}

In the previous section the general expansion of $Y_{1n'}$ is shown. As presented in \cite{oberrath2016}, the author determines an explicit expansion for the MRP with finite dimension
\begin{equation}
Y_{1n'}
=\vec e_1^{T}\cdot\left(i\omega\ul{\op I}-\ul{\op{T}}_{\,V}-\ul{\op{T}}_{\,S}\right)^{-1}\cdot\vec e_{n'}\ .
\label{responseExpSimp}
\end{equation}
The finite expansion is determined by the identity matrix $\ul{\op I}$, the collision matrix $\ul{\op{T}}_{\,S}$, and the Vlasov matrix $\ul{\op{T}}_{\,V}$. $\vec e_{1/2}$ are the explicitly expanded excitation vectors. 

To determine the identity matrix, a complete orthonormal basis is needed. An appropriate basis function in velocity space is given by
\begin{equation}
g_{v}^{\kappa\lambda\mu}(v,\chi,\psi)
=\pi^{\frac{1}{4}}\Lambda_{\kappa}^\lambda(v)\bar P_\lambda^\mu (\cos(\chi))e^{i\mu\psi}\label{BasisKetv}\ .
\end{equation}
The complex exponential function $e^{i\mu\psi}$ is orthogonal and complete on $\psi\in[0,2\pi]$. Its expansion index $\mu\in[-\hat\mu,\hat\mu]$ is restricted to a certain interval. One limit can be defined by $\hat\mu=\lambda$ (A further limit of $\hat\mu$ will be given below.), where $\lambda\in\mathbb N_0$ is the expansion index of $\bar P_\lambda^\mu (\cos(\chi))$. They are the normalized associated Legendre polynomials, which are orthonormal and complete on the interval $\chi\in[0,\pi]$
\begin{equation}
\bar P_\lambda^\mu (\chi)
=\sqrt{\frac{2\lambda+1}{2}\frac{(\lambda-\mu)!}{(\lambda+\mu)!}}P_\lambda^\mu (\cos(\chi))\ .
\end{equation}
 $\kappa\in\mathbb N_0$ is the expansion index of $\Lambda_{\kappa}^\lambda(v)$, which are based on the generalized Laguerre polynomials $L_\kappa^{\lambda+\frac{1}{2}}\left(\frac{1}{2}v^2\right)$. They are an adequate choice on the interval $v\in[0,\infty]$, because the weighting function $w$ has an exponential part in the absolute value of the velocity $v$. By means of an additional factor $L_\kappa^{\lambda+\frac{1}{2}}\left(\frac{1}{2}v^2\right)$ become the orthonormal functions   
\begin{equation}
\Lambda_{\kappa}^{\lambda}(v)
=\sqrt{\frac{\kappa !}{\Gamma(\kappa+\lambda+\frac{3}{2})}}
 \left(\frac{v^2}{2}\right)^{\frac{\lambda}{2}} 
      L_\kappa^{\lambda+\frac{1}{2}}\left(\frac{v^2}{2}\right)\ .
\end{equation}

The scalar product in \eqref{ScalarProduct} is given by two different parts, which makes it difficult to determine an orthogonal function in physical space. As in \cite{oberrath2018}, the author focuses on the first part $\braket{g'}{g}_{\set P}$, which allows to define
\begin{equation}
g_{r}^{kl\mu}(r,\vartheta)=g^{k}(r)\,\Theta_l^\mu(\vartheta)
\label{BasisKetr}\ .
\end{equation}
$\Theta_l^\mu(\vartheta)$ is an orthonormal and complete function on the interval of the polar angle
\begin{equation}
\Theta_l^\mu(\vartheta)=\sqrt{\frac{(2l+1)\,\Gamma(l+1)\, l!}{2\,\Gamma(l-\mu+1)\,\Gamma(l+\mu+1)}}
(1-\cos(\vartheta))^{-\frac{\mu}{2}}(1+\cos(\vartheta))^{\frac{\mu}{2}} 
P_l^{(-\mu,\mu)}(\cos(\vartheta))\ ,
\end{equation} 
which is also an eigenfunction of the Vlasov operator.  
Here, $P_l^{(-\mu,\mu)}(\cos(\vartheta))$ are the Jacobi polynomials with their expansion index $l\in\N_0$. By means of the normalization coefficient one can identify $|\mu|\leq l$, which yields $\hat\mu=\min(l,\lambda)$. Finally, an appropriate basis function on the interval of the radial distance can be defined as
\begin{equation}
g^{k}(r)=\frac{\sin\left(k\pi\frac{r-R}{R_\infty-R}\right)}{\sqrt{2\pi\,e^{\bar\Phi(r)}\,(R_\infty-R)}\, r}
\end{equation}
with $k\in\N$. It is orthonormal and complete on the interval $[R,R_\infty]$ and fulfills the boundary conditions $g^{k}(R)=g^{k}(R_\infty)=0$. 

In summary,  
\begin{equation}
g_{kl}^{\kappa\lambda\mu}(r,\vartheta,v,\chi,\psi)
=g_r^{kl\mu}(r,\vartheta)g_v^{\kappa\lambda\mu}(v,\chi,\psi)
\label{BasisKet}
\end{equation} 
is an orthonormal and complete basis function based on the plasma part of the scalar product $\braket{g'}{g}_{\set P}$. Indeed, this basis function is not orthonormal on the complete scalar product \eqref{ScalarProduct}, but a non-diagonal basis matrix $\ul{\op B}$ can be determine and diagonalized afterwords. Therefore, the complete scalar product of two basis functions has to be computed, which requires the derivative of the inner potential. $\op\phi$ follows Poisson's equation \eqref{PoissonMRP} in which the expanded distribution 
\begin{equation}
g(r,\vartheta,v,\chi ,\psi , \omega)=
\sum_{k=1}^\infty 
\sum_{l=0}^\infty 
\sum_{\kappa=0}^\infty 
\sum_{\lambda=0}^\infty 
\sum_{\mu=-\hat\mu}^{\hat\mu} \tilde g_{kl}^{\kappa\lambda\mu}(\omega) \,
g_{kl}^{\kappa\lambda\mu}(r,\vartheta,v,\chi,\psi)
\end{equation} 
is entered. Expanding also $\op\Phi$ as 
\begin{equation}
\op\Phi(r,\vartheta, \omega)=
\sum_{k=1}^\infty 
\sum_{l=0}^\infty 
\tilde {\op\Phi}_{kl}(\omega) \,
\op\Phi_{k}(r)\Theta_l^0(\vartheta)
=\sum_{k=1}^\infty 
 \sum_{l=0}^\infty 
 \tilde {\op\Phi}_{kl}(\omega) \,
 \op\Phi_{k}(r)\bar P_l(\vartheta)
\end{equation} 
allows to reduce the Poisson equation only to the radial dependence. It is important to note, that $\Theta_l^0(\vartheta)$ collapses to the Legendre polynomial $\bar P_l^0(\vartheta)=\bar P_l(\vartheta)$, which can be utilized to simplify the expansion: the main result from MRP's fluid model with an expansion in Legendre polynomials in the polar angle $\vartheta$ is, that its resonance behavior is dominated by the dipole mode \cite{lapke2013}. In the limit from low pressure to higher pressure the resonance behavior has to be equal and thus, the dipole mode has to be dominant in the kinetic description, too. Due to that, the author takes only this mode into account and sets $l=1$.

Based on the simplified expansion, the radial dependent derivative of the inner potential is given by 
\begin{equation}
\frac{\p\op\Phi_k}{\p r}
=\frac{\delta _{\kappa 0}\, \delta_{\lambda 0}}{r^2}\left\{\begin{matrix}
A_k^{(\set D)} & r\in\set D\\[1ex]
\ds A_k^{(\set P)}
   +\int_R^{r} {r''}^2 \, e^{\bar\Phi(r)}\, g_r^{k} \, dr'' \, dr'  & r\in\set P
\end{matrix}\right.
\ .
\label{SimpGradPotMRP}
\end{equation} 
($\delta _{\kappa 0}$, $\delta_{\lambda 0}$, and $\delta_{\mu 0}$ are Kronecker deltas. The derivation is presented in the Appendix \ref{sec:Pot}). From equation \eqref{SimpGradPotMRP} one can see, that the inner potential $\op\Phi_k$ of a basis function is zero for all $\kappa\neq0$, $\lambda\neq0$, and $\mu\neq0$. This result can be transferred  to the inner potential $\op\Phi_{k'}$ if $\kappa'\neq0$, $\lambda'\neq0$, and $\mu'\neq0$ and simplifies the second part of the scalar product to 
\begin{equation}
\braket{g_{k'1}^{\kappa'\lambda'\mu'}}{g_{k1}^{\kappa\lambda\mu}}_{\set V}
=2\pi\int_{R-d}^{R_\infty}\varepsilon\left(\frac{\p \mathsf\Phi_{k'}}{\p r}\frac{\p \mathsf\Phi_k}{\p r} r^2+2\mathsf\Phi_{k'}\mathsf\Phi_{k}\right)\,dr\,\delta_{\kappa 0}\delta_{\lambda 0}\delta_{\mu 0}\delta_{\kappa' 0}\delta_{\lambda' 0}\delta_{\mu' 0}=\op B_{kk'}^{(000)}\ .
\end{equation} 
$\op B_{kk'}^{(000)}$ is not zero only if $\lambda=\lambda'=\kappa=\kappa'=\mu=\mu'=0$ and leads to the final result of the full scalar product of two basis functions, which can be separated into two different parts
\begin{eqnarray}
\braket{g_{k'1}^{000}}{g_{k1}^{000}} & = & 
\op B_{kk'}^{(000)}+\delta_{kk'}\label{basisB0}\ ,\\
\braket{g_{k'1}^{\kappa'\lambda'\mu'}}{g_{k1}^{\kappa\lambda\mu}} & = & \delta_{kk'}\delta_{\kappa\kappa'}\delta_{\lambda\lambda'}\delta_{\mu\mu'}\ .
\end{eqnarray}
This result of the scalar product defines the elements of the basis matrix $\ul{\op B}$, which is a block diagonal matrix. It is almost diagonal, except the first matrix $\ul{\op B}^{(000)}$ if $\lambda=\lambda'=\kappa=\kappa'=\mu=\mu'=0$. However, it can be diagonalized with a rotation matrix $\ul{\op C}$ to find the diagonal matrix $\ul{\op D}^{(000)}=\ul{\op C}\,\ul{\op B}^{(000)}\, \ul{\op C}^T$. Normalizing $\ul{\op D}^{(000)}$ with its inverse leads to the identity matrix $\ul{\op I}^{(000)}=\ul{\op D}^{(000)}\, {\ul{\op D}^{(000)}}^{-1}$. Then, $\ul{\op B}$ becomes the pure identity matrix $\ul{\op I}$ and is the equivalent of an expansion with an orthonormal and complete system of basis functions.  

Since an appropriate system of basis functions is defined, the operators have to be expanded with this system. The simpler operator is the collision operator, which will be expanded first. Applying it to the basis function and computing the scalar product leads to the matrix elements of the collision matrix $\ul{\op T}_{\,S}$
\begin{equation}
\langle g_{k'1}^{\kappa'\lambda'\mu'} | \op T_S g_{k1}^{\kappa\lambda\mu}\rangle
=\langle g_{k'1}^{\kappa'\lambda'\mu'} | \op T_S g_{k1}^{\kappa\lambda\mu}\rangle_{\set P}
=\, \nu_0 \,\left(\delta_{\lambda 0}\,\delta_{\lambda' 0}\delta_{\mu 0}\,\delta_{\mu' 0}-\delta_{\lambda\lambda'}\delta_{\mu\mu'}\right) 
  \,\delta_{\kappa\kappa'}\delta_{kk'} \label{MeTs}\ .
\end{equation}
$\ul{\op T}_{\,S}$ is a pure diagonal matrix and in addition with zero elements on the main diagonal if $\lambda=\lambda'=0$. Due to that, no correction by the diagonalization is needed. 

The derivation of the Vlasov matrix $\ul{\op T}_{\,V}$ is more complicated. Its elements are determined by the scalar product between the basis functions and the Vlasov operator $\langle g_{k'1}^{\kappa'\lambda'\mu'} | \op T_V g_{k1}^{\kappa\lambda\mu}\rangle$. As shown in \cite{oberrath2016}, $\ul{\op T}_{\,V}$ is an anti-symmetric block matrix. Due to that only inner block matrices with the indices $\kappa=\kappa'=\lambda=0$, $\lambda'=1$ and $\kappa=\kappa'=\lambda'=0$, $\lambda=1$ have to be multiplied with the rotation matrices $\ul{\op C}$ or $\ul{\op C}^T$ to get the correct expanded Vlasov matrix. Detailed calculations to the Vlasov matrix can be found in the appendix \ref{sec:VlasovMatrix}. 

Finally, the excitation vectors $\vec e_{1/2}$ have to be determined. They include the electrode functions $\psi_{1/2}$, which are solved in appendix \ref{sec:char}. Due to the anti-symmetric excitation of the probe, only the dipole mode has to be taken into account and the elements of the excitation vectors are then defined by
\begin{eqnarray}
\braket{g_{k'1}^{\kappa'\lambda'\mu'}}{e_{1/2}}
& = & 2\pi\int_R^{R_\infty} r^2 \, e^{\bar\Phi(r)}\, g^{k'}\, \frac{\p\psi_{1/2}}{\p r} \,dr
       \,\delta_{\mu' 0}\,\delta_{\lambda' 1}\,\delta_{\kappa' 0}\nonumber\\[1ex]
&   &-2\pi\int_R^{R_\infty} r \, e^{\bar\Phi(r)}\, g^{k'}\, \psi_{1/2} \,dr
       \left(\delta_{\mu' 1}+\delta_{-1 \mu'}\right)\delta_{\lambda' 1}\,\delta_{\kappa' 0} 
       \nonumber\\[1ex]
& = &  \pm\left(e_{k'}^0\,\delta_{\mu' 0}-e_{k'}^1\left(\delta_{\mu' 1}+\delta_{-1 \mu'}\right)\right)
       \,\delta_{\lambda' 1}\,\delta_{\kappa' 0}\ .
\end{eqnarray}
By means of the Kronecker Deltas it is obvious, that $\vec e_{1/2}$ have non-vanishing elements only for $\kappa'=0$ and $\lambda'=1$. Thus, the full vectors can be written as
\begin{equation}
\vec e_{1/2}=\pm\left(\vec 0\, ,\, {\ul{\op D}^{(000)}}^{-\frac{1}{2}}\,\ul{\op C}\,\vec e_{k'}\, ,\, \vec 0\, ,\, \dots \right)^T\ .
\end{equation}

Now, all elements of the expanded admittance $Y_{1n'}$ in \eqref{responseExpSimp} are determined and the expanded current at electrode $\set E_1$ can be defined. The dipole excitation of the probe with $U_1=-U_2=U$ yields also anti-symmetric excitation vectors $\vec e_{1}=-\vec e_{2}$, which can be utilized to simplify the current
\begin{equation}
i_1=\sum_{n'=1}^2 \vec e_1^{T}\cdot
                  \left(i\omega\ul{\op I}-\ul{\op{T}}_{\,V}-\ul{\op{T}}_{\,S}\right)^{-1}
                  \cdot\vec e_{n'}\, U_{n'}
   =2\vec e_1^{T}\cdot\left(i\omega\ul{\op I}-\ul{\op{T}}_{\,V}-\ul{\op{T}}_{\,S}\right)^{-1}
                 \cdot\vec e_1\,U=Y_{\rm MRP}\,U.
\label{currentApprox}
\end{equation}
$Y_{\rm MRP}$ is then the explicitly expanded admittance of the MRP and can be used to compute different spectra.

\section{Spectra of the spherical multipole resonance probe}

Within the last section an explicit expansion of the inner admittance of the idealized spherical MRP is derived and can be used to compute approximated spectra. To compare the first calculated kinetic spectra of the MRP all parameters are taken from Buckley \cite{buckley1967} and the spectra will be compared to the kinetic spectra of the spherical IP \cite{oberrath2018}. 

These spectra are calculated for a MRP without dielectric ($d=0$, $\varepsilon_D=1$) and a probe radius of $R=5.15\,\lambda_D$. The equilibrium potential $\bar\Phi(r)$ of a spherical electrode in a plasma presented by Bernstein and Rabinowitz \cite{bernstein1959} is applied in the calculations. The collision frequency is varied as $\nu_0\in\{0.05 ,0.15, 0.25\}\, \omega_{\rm p}$. The distance to the outer grounded surface is chosen to be $R_\infty=150\lambda_D$, where also the plasma frequency is normalized to $\omega_{\rm p}$. 
\begin{figure}[h!]
\centering
\includegraphics[height=7cm]{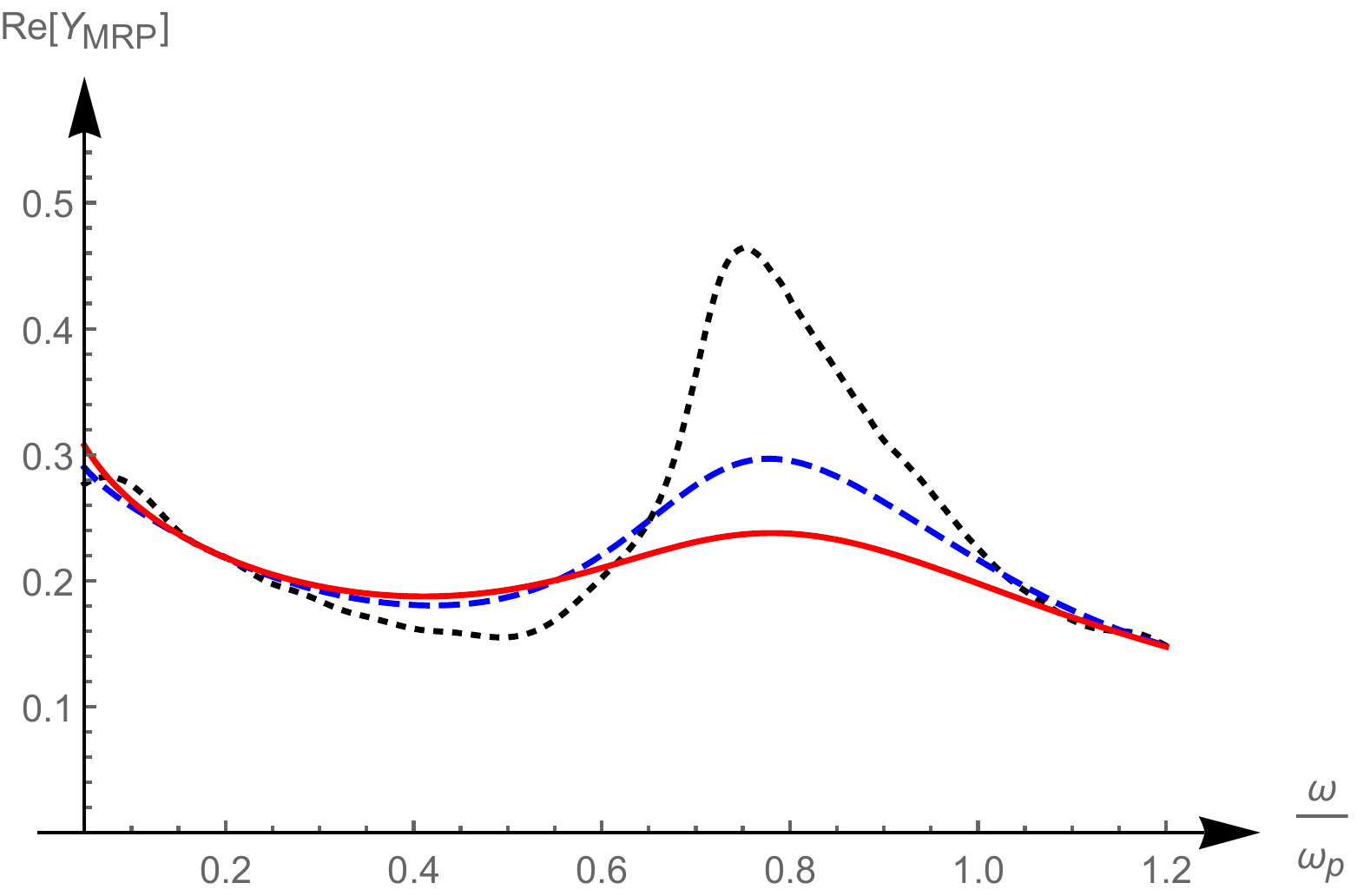}
\vspace{-3mm}
\caption{Normalized real part of the admittance of the MRP for $\kappa_{\rm max}=20$, $\lambda_{\rm max}=15$, $k_{\rm max}=500$ and different collision frequencies $\nu_0\omega_{\rm p}^{-1}$: 0.05 (dotted), \textcolor{blue}{0.15~(dashed)}, and \textcolor{red}{0.25~(bold)}.}
\label{YMRP}
\end{figure}

\begin{figure}[h!]
\centering
\includegraphics[height=4.65cm]{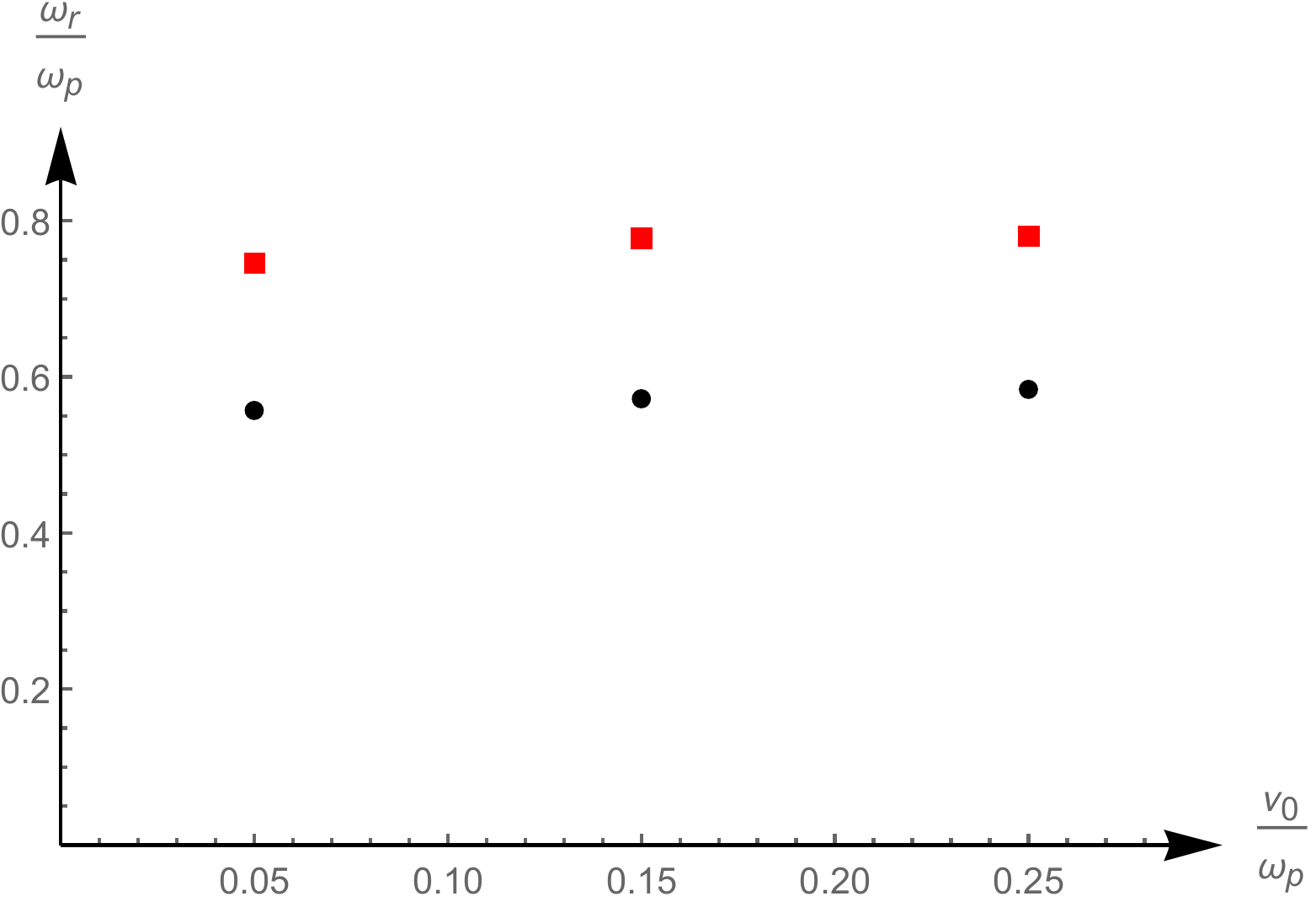}
\hspace{8mm}
\includegraphics[height=4.65cm]{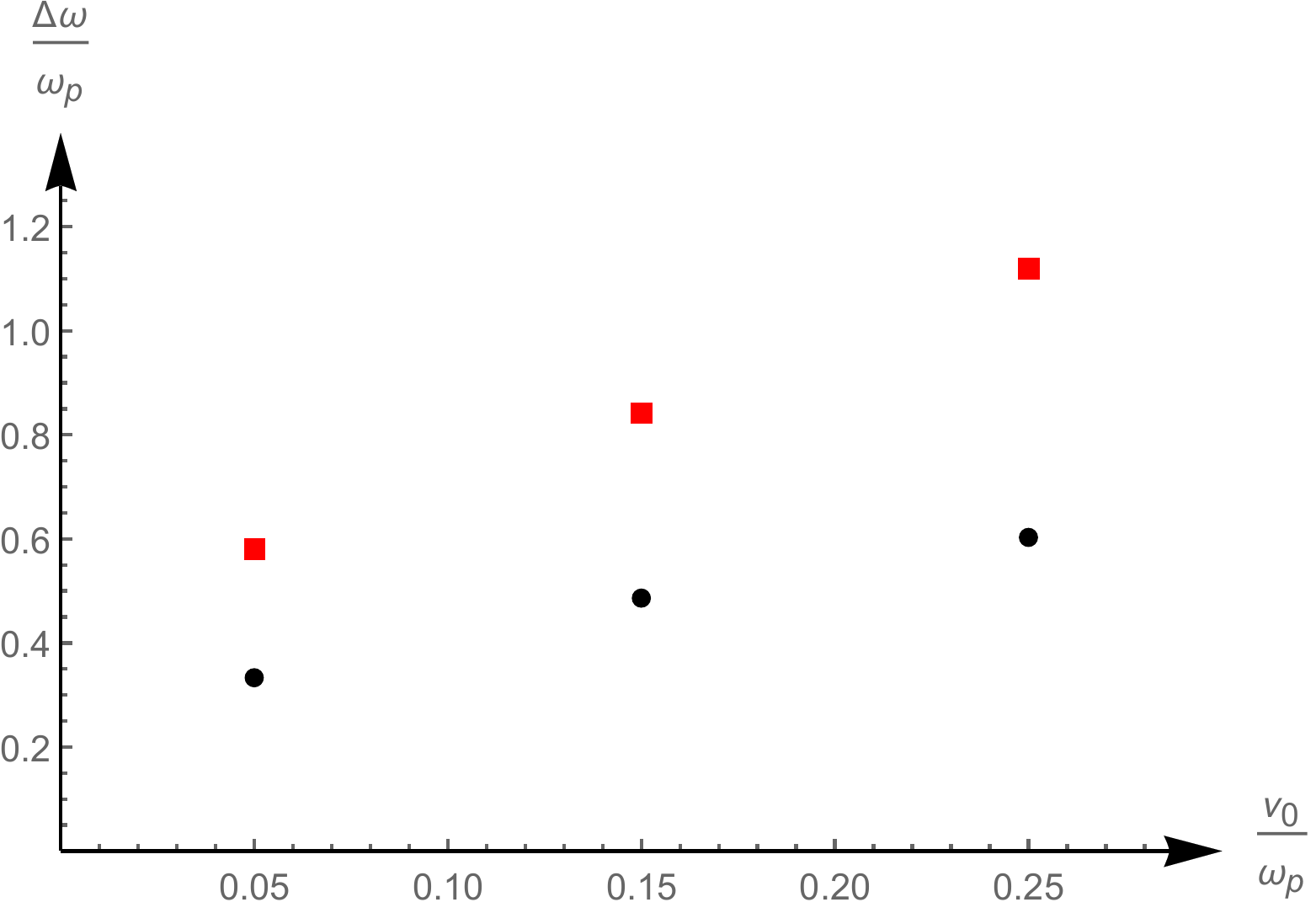}
\caption{Comparison of the resonance frequencies (left) and the half widths (right) of the \textcolor{red}{MRP (squares)} and the IP (dots) for three different collision frequencies $\nu_0\omega_{\rm p}^{-1}$: $0.05$, $0.15$, and $0.25$. 
}
\label{ResF_HWB_MRP}
\end{figure}

In fig. \ref{YMRP} the real part of the admittance $Y_{\rm MRP}$ is depicted for the maximum expansion indices in velocity space $\kappa_{\rm max}=20$, $\lambda_{\rm max}=15$ and a maximum expansion index $k_{\rm max}=500$ in physical space. From these almost converged spectra the resonance frequencies can be determined as $\omega_{\rm r}\in\{0.751,0.777,0.781\}\, \omega_{\rm p}$, which are about $0.2\,\omega_{\rm p}$ larger than the resonance frequencies from the IP, as depicted in fig. \ref{ResF_HWB_MRP} (left). Such a difference of the resonance frequencies of the monopole and dipole mode is expected and also present in spectra determined by a fluid model \cite{oberrath2014b}. The half widths $\Delta\omega$ of the resonance peaks in the spectra of the MRP are broader than the half widths of the IP spectra and increase for higher collision frequencies (see fig. \ref{ResF_HWB_MRP}, right). This indicates, that higher modes are stronger influenced by kinetic effects as lower modes. Here, the half widths of the MRP are determined by the right part of the resonance peak and then multiplied by the factor of two, due to the asymmetric peak shapes.  

The half widths in the spectra represent the damping of the probe-plasma system. They are larger than determined by fluid models, which is caused by kinetic effects. Assuming $\Delta\omega=\nu_0+\nu_{\rm kin}$, the kinetic damping can be determined as $\nu_{\rm kin}\in\{0.531, 0.692, 0.87\}\, \omega_{\rm p}$.      

\section{Conclusion}

Within this manuscript a kinetic model of the spherical multipole resonance probe is presented, where its dynamical interaction with the surrounding plasma is given by the inner admittance of the probe-plasma system. This admittance is determined by the resolvent of the dynamical operator $\op T_V + \op T_S$, which has to be expanded to allow for the computation of the corresponding spectra. The expanded inner admittance of the MRP is derived by means of a complete basis in its spherical geometry and leads to the matrix representation of the dynamical operator. Truncating the expansion allows to approximate the inner admittance and thus to analyze the kinetic damping within its spectra.  

To compare the approximated spectra of the MRP, the parameters in the calculations are taken from former computations of spectra of the IP \cite{oberrath2018} for the three different collision frequencies $\nu_0\in\{0.05,0.15,0.25\}\,\omega_{\rm p}$. The resonance frequencies of the MRP, which excites a dipole mode, are about $0.2\,\omega_{\rm p}$ higher as the ones from the IP, which excites a monopole mode. Such a difference can also be observed in spectra determined by a fluid model \cite{oberrath2014b}. 

In addition the half widths of the resonance peaks are determined. It is shown, that the half widths of the MRP spectra are broader than the ones from the IP. This indicates, that higher resonance modes are stronger effected by kinetic effects as the monopole mode, which can be explained by the kinetic loss mechanism as described in ref. \cite{oberrath2014}: the probe produces kinetic free energy, which is transported through the plasma and escapes at a large distance to the probe, where the probe can not detect it anymore. This loss of kinetic free energy is recorded in the spectrum of the probe as damping. In other words, particles in the detection region gain energy by the probe and leave this region. Due to the fact, that the electric field of the dipole mode decreases with $r^{-3}$, the detection region is smaller than the detection region of the monopole mode, where the field decreases with $r^{-2}$. It can be assumed, that the kinetic damping $\nu_{\rm kin}$ is proportional to the fraction of the thermal velocity of the electrons $v_{\rm th}$ and anti-proportional to the dimension of the detection region $R_{\rm det}$ $\left(\nu_{\rm kin}\propto\frac{v_{\rm th}}{R_{\rm det}}\right)$. Thus, more particles, which gain energy by the probe, can easier or faster leave the detection region, which is recorded in the spectrum of the MRP as stronger kinetic damping.

In summary it is shown that the approximated spectra of the MRP based on the functional analytic approach show resonance frequencies and half widths as expected from former results of the impedance probe. The stronger kinetic damping is explainable by the decreased influence range of the dipole mode. In further works the parameter of the real MRP will be applied, to compare the results with measurements and derive relations between the resonance parameter and electron density and temperature.

\appendix

\section{Potential of a basis function}\label{sec:Pot}

Entering the basis function \eqref{BasisKet} into Poisson's equation \eqref{PoissonMRP} the integrals over the velocity space can be solved which yields 
\begin{equation}
\frac{\p}{\p r}\left(r^2\frac{\p\op\Phi_k}{\p r}\right)-2\op\Phi_k
=\delta_{\kappa 0}\delta_{\lambda 0}\delta_{\mu 0}\left\{\begin{matrix}
0 & , & r\in\set D\\[1ex]
\ds \, r^2 e^{\bar\Phi(r)}\, g_r^{k}(r)  & , & r\in\set P
\end{matrix}\right.
\ .
\label{SimpPoissonMRP}
\end{equation} 

The potentials for the different regions (plasma $\set P$ and dielectric $\set D$) can be solved by integration and using the boundary conditions $\op\Phi_{k}^{(\set P)}(R_\infty)=0$ and $\op\Phi_{k}^{(\set D)}(R-d)=0$: 
\begin{eqnarray}
\op\Phi_{k}^{(\set P)}(r)
&=&A_k^{(\set P)} \left(r-\frac{R_\infty^3}{r^2}\right)
   \delta _{\kappa 0}\, \delta_{\lambda 0}\, \delta_{\mu 0}\nonumber\\[1ex]
& &+\frac{1}{3}\int_R^{r} \left(r-\frac{{r'}^3}{r^2}\right) \, e^{\bar\Phi(r')}\, g^{k} \, dr'
   \,\delta _{\kappa 0}\, \delta_{\lambda 0}\, \delta_{\mu 0}\\[1ex]
& &+\frac{1}{3r^2}\int_R^{R_\infty} \left({r'}^3-R_\infty^3\right)\, e^{\bar\Phi(r')}\, g^{k}\, dr'
    \,\delta _{\kappa 0}\, \delta_{\lambda 0}\, \delta_{\mu 0}\nonumber \ ,\\[1ex]
\op\Phi_{k}^{(\set D)}(r)
&=&A_{k}^{(\set D)} \left(r-\frac{(R-d)^3}{r^2}\right)
   \delta _{\kappa 0}\, \delta_{\lambda 0}\, \delta_{\mu 0} \ .  
\end{eqnarray}
The constants $A_k^{(\set P)}$ and $A_k^{(\set D)}$ are determined by the transition conditions
\begin{eqnarray}
\op\Phi_{k}^{(\set D)}(R) & = & \op\Phi_{k}^{(\set P)}(R)\ , \\[1ex]
\epsi_D \left.\frac{\p\op\Phi_{k}^{(\set D)}}{\p r}\right|_R
& = &  \left.\frac{\p\op\Phi_{k}^{(\set P)}}{\p r}\right|_R \ .
\end{eqnarray}
In the scalar product also the derivative of the potential is needed and can be written as 
\begin{eqnarray}
\frac{\p}{\p r}\op\Phi_{k}^{(\set P)}(r)
&=&A_k^{(\set P)} \left(1+\frac{2R_\infty^3}{r^3}\right)
   \delta _{\kappa 0}\, \delta_{\lambda 0}\, \delta_{\mu 0}\nonumber\\[1ex]
& &+\frac{1}{3}\int_R^{r} \left(1+\frac{2{r'}^3}{r^3}\right) \, e^{\bar\Phi(r')}\, g^{k} \, dr'
   \,\delta _{\kappa 0}\, \delta_{\lambda 0}\, \delta_{\mu 0}\\[1ex]
& &-\frac{2}{3r^3}\int_R^{R_\infty} \left({r'}^3-R_\infty^3\right)\, e^{\bar\Phi(r')}\, g^{k}\, dr'
    \,\delta _{\kappa 0}\, \delta_{\lambda 0}\, \delta_{\mu 0}\nonumber \ ,\\[1ex]
\frac{\p}{\p r}\op\Phi_{k}^{(\set D)}(r)
&=&A_{k}^{(\set D)} \left(1+\frac{2(R-d)^3}{r^3}\right)
   \delta _{\kappa 0}\, \delta_{\lambda 0}\, \delta_{\mu 0} \ .  
\end{eqnarray}

\section{Matrix elements of the Vlasov-Operator}\label{sec:VlasovMatrix}

The Vlasov operator is defined in \eqref{VlasovOp}. Applied to the basis function $g_{k1}^{\kappa\lambda\mu}$ yields
\begin{align}
\op T_V g_{k1}^{\kappa\lambda\mu} 
= & v\left(\cos(\chi)\frac{\p}{\p r}
    +\frac{\sin(\chi)\cos(\psi)}{r}\frac{\p}{\p\vartheta}\right)
     \left(\op\Phi - g_{k1}^{\kappa\lambda\mu}\right) 
    + \frac{\p \bar\Phi}{\p r}
      \left(\frac{\sin(\chi)}{v}\frac{\p g_{k1}^{\kappa\lambda\mu}}{\p \chi}
            -\cos(\chi)\frac{\p g_{k1}^{\kappa\lambda\mu}}{\p v}\right)\nonumber\\
&   +\sin(\chi)\frac{v}{r}\left(\frac{\p g_{k1}^{\kappa\lambda\mu}}{\p\chi}
    +\cot(\vartheta)\sin(\psi)\frac{\p g_{k1}^{\kappa\lambda\mu}}{\p\psi}\right)\ .  
\end{align}
In the scalar product the gradient of the potential $\op\Phi^{(\op T_V)}$ is needed, which is meant as the potential produced by the Vlasov operator applied to the basis function. In \cite{oberrath2016} is shown that this gradient is given by the electron particle flux within the plasma $\set P$ and vanishes within the dielectric $\set D$. In the geometry of the MRP one finds
\begin{eqnarray}
\frac{\p\op\Phi_k^{(\op T_V)}}{\p r}
& = & -\frac{1}{(2\pi)^{\frac{3}{2}}}\,g_r^{k1\mu}\, e^{\bar\Phi(r)}
       \int_{0}^{2\pi}\int_{0}^\pi\int_0^{\infty } e^{-\frac{v^2}{2}}    
       g_{v}^{\kappa\lambda\mu}\, v^3 \sin(\chi)\cos(\chi)\,dv\,d\chi\,d\psi\ ,\\[1ex]
\frac{\p\op\Phi_k^{(\op T_V)}}{\p\vartheta} 
& = & -\frac{1}{(2\pi)^{\frac{3}{2}}}\,g_r^{k1\mu}\, e^{\bar\Phi(r)}
       \int_{0}^{2\pi}\int_{0}^\pi\int_0^{\infty } e^{-\frac{v^2}{2}}\, 
       g_{v}^{\kappa\lambda\mu}v^3 \sin^2(\chi)\cos(\psi)\,dv\,d\chi\,d\psi\ .       
\end{eqnarray}
Due to that, the elements of the Vlasov matrix are given by
\begin{align}
  &\langle g_{k'1}^{\kappa'\lambda'\mu'} | \op T_V | g_{k1}^{\kappa\lambda\mu}\rangle\label{SPTv}\\[1ex]
= &\,\frac{{\op V}_{kk'}^{(1)}}{(2\pi)^{\frac{3}{2}}}\left[
   \int_{0}^\pi\int_{0}^{2\pi}\int_{0}^\pi\int_0^{\infty} 
    e^{-\frac{v^2}{2}}\, \left(g_{v}^{\kappa'\lambda'\mu'}\right)^*\, \Theta_{1}^{\mu'}\,
    \Theta_{1}^{\mu}\,\frac{\p g_{v}^{\kappa\lambda\mu}}{\p\chi}\, v\sin^2(\chi)\sin(\vartheta)
    \,dv\,d\chi\,d\psi\,d\vartheta\right.\nonumber\\[1ex]
  &\qquad\qquad-\left.\int_{0}^\pi\int_{0}^{2\pi}\int_{0}^\pi\int_0^{\infty} 
    e^{-\frac{v^2}{2}}\, \left(g_{v}^{\kappa'\lambda'\mu'}\right)^*\, \Theta_{1}^{\mu'}\,
    \Theta_{1}^{\mu}\,\frac{\p g_{v}^{\kappa\lambda\mu}}{\p v}\, \frac{v^2}{2}\sin(2\chi)
    \sin(\vartheta)\,dv\,d\chi\,d\psi\,d\vartheta\right]\nonumber\\[1ex] 
  &+\frac{{\op V}_{kk'}^{(2)}}{(2\pi)^{\frac{3}{2}}}\int_{0}^\pi\int_{0}^{2\pi}\int_{0}^\pi
    \int_0^{\infty } e^{-\frac{v^2}{2}}\, \left(g_{v}^{\kappa'\lambda'\mu'}\right)^*
    \, g_{v}^{\kappa\lambda\mu}\, \Theta_{1}^{\mu'}\, \Theta_{1}^{\mu}\, \frac{v^3}{2}\sin(2\chi)
    \sin(\vartheta)\,dv\,d\chi\,d\psi\,d\vartheta\nonumber\\[1ex]
  &+\frac{ {\op V}_{kk'}^{(3)}}{(2\pi)^{\frac{3}{2}}}\left[\int_{0}^\pi\int_{0}^{2\pi}\int_{0}^\pi
    \int_0^{\infty } e^{-\frac{v^2}{2}}\, \left(g_{v}^{\kappa'\lambda'\mu'}\right)^*\,
    \Theta_{1}^{\mu'}\, \Theta_{1}^{\mu}\,\frac{\p g_{v}^{\kappa\lambda\mu}}{\p\chi}\, v^3 
    \sin^2(\chi)\sin(\vartheta) \,dv\,d\chi\,d\psi\,d\vartheta\right.\nonumber\\[1ex]
  & \qquad\qquad+\int_{0}^\pi\int_{0}^{2\pi}\int_{0}^\pi\int_0^{\infty }
    e^{-\frac{v^2}{2}}\, \left(g_{v}^{\kappa'\lambda'\mu'}\right)^*\,\Theta_{1}^{\mu'}\, 
    \Theta_{1}^{\mu} \,\frac{\p g_{v}^{\kappa\lambda\mu}}{\p\psi}\, v^3 \sin^2(\chi)\sin(\psi)
    \cos(\vartheta)\,dv\,d\chi\,d\psi\,d\vartheta\nonumber\\[1ex] 
   & \left.\qquad\qquad-\int_{0}^\pi\int_{0}^{2\pi}\int_{0}^\pi\int_0^{\infty }
    e^{-\frac{v^2}{2}}\,\left(g_{v}^{\kappa'\lambda'\mu'}\right)^*\,\Theta_{1}^{\mu'}\, 
    g_{v}^{\kappa\lambda\mu}\,\frac{\p \Theta_{1}^{\mu}}{\p\vartheta}\, v^3 \sin^2(\chi)\cos(\psi)
    \sin(\vartheta)\,dv\,d\chi\,d\psi\,d\vartheta\right]\nonumber\\[1ex] 
  &+\frac{{\op V}_{kk'}^{(4)}}{(2\pi)^{\frac{3}{2}}}\int_{0}^\pi\int_{0}^{2\pi}\int_{0}^\pi
    \int_0^{\infty } e^{-\frac{v^2}{2}}\,\left(g_{v}^{\kappa'\lambda'\mu'}\right)^*\,
    \Theta_{1}^{\mu'}\, \bar P_{1}\, \frac{v^3}{2}\sin(2\chi) \sin(\vartheta)\,dv\,d\chi\,d\psi
    \,d\vartheta \nonumber\\[1ex] 
  &+\frac{{\op V}_{kk'}^{(5)}}{(2\pi)^{\frac{3}{2}}}\int_{0}^\pi\int_{0}^{2\pi}\int_{0}^\pi
    \int_0^{\infty } e^{-\frac{v^2}{2}}\, \bar P_{1} \,\Theta_{1}^{\mu}\, g_{v}^{\kappa\lambda\mu}
    \,\frac{v^3}{2}\sin(2\chi)\sin(\vartheta)\,dv\,d\chi\,d\psi\,d\vartheta\nonumber\\[1ex]  
  &+\frac{{\op V}_{kk'}^{(6)}}{(2\pi)^{\frac{3}{2}}}\int_{0}^\pi\int_{0}^{2\pi}\int_{0}^\pi
    \int_0^{\infty } e^{-\frac{v^2}{2}}\, \left(g_{v}^{\kappa'\lambda'\mu'}\right)^*\,
    \Theta_{1}^{\mu'}\, \frac{\p\bar P_{1}}{\p\vartheta}\, v^3 \sin^2(\chi)\cos(\psi)
    \sin(\vartheta)\,dv\,d\chi\,d\psi \,d\vartheta \nonumber\\[1ex] 
  &+\frac{{\op V}_{kk'}^{(7)}}{(2\pi)^{\frac{3}{2}}}\int_{0}^\pi\int_{0}^{2\pi}\int_{0}^\pi
    \int_0^{\infty } e^{-\frac{v^2}{2}}\, \frac{\p\bar P_{1}}{\p\vartheta} \,\Theta_{1}^{\mu}\, 
    g_{v}^{\kappa\lambda\mu}v^3 \sin^2(\chi)\cos(\psi)\sin(\vartheta)\,dv\,d\chi\,d\psi
    \,d\vartheta\nonumber
\end{align}
with
\begin{eqnarray}
{\op V}_{kk'}^{(1)} 
& = & 2\pi\int_R^{R_\infty} r^2\, g^{k'}\,\frac{\p e^{\bar\Phi(r)}}{\p r}\,g^{k} \, dr\ ,\label{SPTvV1}\\[1ex]
{\op V}_{kk'}^{(2)} 
& = & -2\pi\int_R^{R_\infty} r^2\, g^{k'}\, e^{\bar\Phi(r)} \frac{\p g^{k}}{\p r}\, dr\ ,\\[1ex]
{\op V}_{kk'}^{(3)}
& = & 2\pi\int_R^{R_\infty} r\,g^{k'}\, e^{\bar\Phi(r)}\, g^{k} \, dr\ ,\\[1ex]
{\op V}_{kk'}^{(4)} 
& = & 2\pi\int_R^{R_\infty} r^2\,g^{k'}\, e^{\bar\Phi(r)}\,\frac{\p\op\Phi_{k}^{(\set P)}}{\p r}\, dr=-{\op V}_{k'k}^{(5)}\ ,\\[1ex]
{\op V}_{kk'}^{(5)} 
& = & -2\pi\int_R^{R_\infty} r^2 \,\frac{\p\op\Phi_{k'}^{(\set P)}}{\p r}\, e^{\bar\Phi(r)}\,g^{k}\, dr
\end{eqnarray}
\begin{eqnarray}
{\op V}_{kk'}^{(6)} 
& = & 2\pi\int_R^{R_\infty} r\,g^{k'}\, e^{\bar\Phi(r)}\,\op\Phi_{k}^{(\set P)}\, dr
=-{\op V}_{k'k}^{(7)} \ ,\\[1ex]
{\op V}_{kk'}^{(7)} 
& = & -2\pi\int_R^{R_\infty} r \,\op\Phi_{k'}^{(\set P)}\, e^{\bar\Phi(r)}\,g^{k}\, dr\label{SPTvV7}\ .
\end{eqnarray}
The integrals over the velocity space in \eqref{SPTv} can be solved analytically, but lead to long expressions. The integrals over the physical space in equations \eqref{SPTvV1} to \eqref{SPTvV7} have usually to be solved numerically, depending on the equilibrium potential $\bar\Phi(r)$.

The final Vlasov matrix $\ul{\op T}_{\,V}$ is an anti-symmetric block matrix, where the inner blocks are given by the matrices of the physical space ${\ul{\op V}^{(i)}}$ over the indices $k$ and $k'$. Due to the anti-symmetry, only the block matrices at the positions with the indices $\kappa=\kappa'=\lambda=0$, $\lambda'=1$ and $\kappa=\kappa'=\lambda'=0$, $\lambda=1$ have to be corrected for the complete orthonormal expansion. The correct block matrices at these positions are ${\ul{\op D}^{(000)}}^{-1/2}\,\ul{\op C} \,{\ul{\op V}^{(i)}}$ for $\kappa=\kappa'=\lambda=0$, $\lambda'=1$ and $ {\ul{\op V}^{(i)}}\,\ul{\op C}^T\, {\ul{\op D}^{(000)}}^{-1/2}$ for $\kappa=\kappa'=\lambda'=0$, $\lambda=1$. After this correction the complete Vlasov matrix can be computed as
\begin{equation}
\ul{\op T}_{\,V}=\sum_{i=1}^7 \ul{\op T}_{\,V}^{(i)}\ .
\end{equation}

\section{Solution of the electrode functions}\label{sec:char}

To determine the electrode function, which fulfill the Laplace equation in \eqref{Laplace}, $\psi_{1/2}$ can be expanded in the Legendre polynomials $\bar P_l(\vartheta)$ and yields
\begin{equation}
 \frac{\p}{\p r}\left(r^2\frac{\p}{\p r}\psi_{1/2}^{(\set{D}/\set{P})}\right)
-\frac{l(l+1)}{r^2}\psi_{1/2}^{(\set{D}/\set{P})}=0\ .
\end{equation}
The solution of the radial dependent Laplace equation is given by
\begin{equation}
\psi_{1/2}^{(\set{D}/\set{P})}
=\alpha_{1/2}^{(\set{D}/\set{P})} r^l
+\beta_{1/2}^{(\set{D}/\set{P})}r^{-(l+1)}\ .
\end{equation}
By means of the boundary condition $\psi_{1/2}^{(\set P)}(R_{\infty},\vartheta)=0$ and the transition conditions 
\begin{eqnarray}
\psi_{1/2}^{(\set{P})}(R) & = & \psi_{1/2}^{(\set{D})}(R)\ , \\[1ex]
\frac{\p}{\p r}\psi_{1/2}^{(\set{P})}(R) & = & 
\varepsilon_D\frac{\p}{\p r}\psi_{1/2}^{(\set{D})}(R)\ ,
\end{eqnarray} 
six coefficients can be determined
\begin{eqnarray}
\alpha_{1/2}^{(\set{P})} & = & \frac{(2l+1) \varepsilon_D}{l (\varepsilon_D-1) R^{2l+1}-R_\infty^{2 l+1} (l\varepsilon_D+l+1)}\beta_{1/2}^{(\set{D})}\ ,\\[1ex]
\beta_{1/2}^{(\set{P})} & = & \frac{(2l+1)\varepsilon_D R_\infty^{2l+1}}{R_\infty^{2l+1} (l\varepsilon_D+l+1)-l (\varepsilon_D-1) R^{2l+1}}
\beta_{1/2}^{(\set{D})}\ , \\[1ex]
\alpha_{1/2}^{(\set{D})} & = & \frac{R^{2 l+1} (l\varepsilon_D+l+\varepsilon_D)-(l+1) (\varepsilon_D-1) R_\infty^{2l+1}}{l(\varepsilon_D-1) R^{2l+1}-R_\infty^{2l+1}(l\varepsilon_D+l+1)}
\frac{\beta_{1/2}^{(\set{D})}}{R^{2l+1}}\ .
\end{eqnarray}
The expanded electrode functions can the be written as
\begin{equation}
\psi_{1/2}(r,\vartheta)=\sum_{l=0}^\infty \beta_{1/2}^{(\set{D})}\, 
 \tilde\psi_{1/2}(r)\, \bar P_l(\vartheta)
\end{equation}
Two coefficients are undefined, yet, and have to be solved separately with the boundary conditions 
\begin{eqnarray}
\psi_1^{(\set D)}(R-d,\vartheta) & = & 
\left\{\begin{matrix}
1 & , & \vartheta\in\left[0,\frac{\pi}{2}\right]\\[1ex]
0 & , & \vartheta\in\left(\frac{\pi}{2},\pi\right]
\end{matrix}\right. \ ,\\[1ex]
\psi_2^{(\set D)}(R-d,\vartheta) & = & \left\{\begin{matrix}
0 & , & \vartheta\in\left[0,\frac{\pi}{2}\right]\\[1ex]
1 & , & \vartheta\in\left(\frac{\pi}{2},\pi\right]
\end{matrix}\right.\ .
\end{eqnarray} 
Therefore the orthonormal relation of the Legendre polynomials can be utilized and yields
\begin{align}
\beta_{1/2}^{(\set{D})}=&
\frac{R R_\infty(d-R)}{\sqrt{2}(\varepsilon_D(d-R)(R_\infty-R)-dR_\infty)}
\delta_{l0}\\[1ex]
&\mp\frac{\sqrt{\frac{\pi}{2}}\frac{\sqrt{2l+1}}
 {l\Gamma\left(-\frac{l}{2}\right)\Gamma\left(\frac{l+3}{2}\right)}
 \left(1-\frac{d}{R}\right)^{-(2l+1)} \left(l (\varepsilon_D-1) \left(\frac{R}{R_\infty}\right)^{2l+1}-(l(\varepsilon_D+1)+1)\right)(R-d)^{l+1}}
{\left(\frac{R}{R_\infty}\right)^{2l+1} ((l+1)\varepsilon_D+l)-(l+1) (\varepsilon_D-1)
+\frac{l (\varepsilon_D-1) \left(\frac{R}{R_\infty}\right)^{2l+1}- (l \varepsilon_D+l+1)}{\left(1-\frac{d}{R}\right)^{2l+1}}}\nonumber
\end{align} 

\acknowledgments
The author acknowledges support by the German Research Foundation via the project OB 469/1-1. Gratitude is expressed to J.~Gong, D.-B.~Grys, M.~Lapke, M.~Oberberg, D.~Pohle, C.~Schulz, J.~Runkel, R.~Storch, T.~Styrnoll, S.~Wilczek, T.~Mussenbrock, P.~Awakowicz, T.~Musch, and I.~Rolfes, who are or were part of the MRP-Team at Ruhr University Bochum. Explicit gratitude is expressed to R.P.~Brinkmann for fruitful discussions.


\clearpage

\begin{thebibliography}{10}

 

\bibitem{takayama1960}
K. Takayama, H. Ikegami, and S. Miyazaki, Phys. Rev. Let. {\bf 5}, 238 (1960).


\bibitem{levitskii1963}
S.M. Levitskii and I.P. Shashurin, Sov. Phys. Tech. Phys. {\bf 8}, 319 (1963).

 








\bibitem{harp1964} R. S. Harp, Appl. Phys. Lett. {\bf 4}, 186 (1964).

\bibitem{fejer1964} J.A. Fejer, Radio Sci. {\bf 68D}, 1171 (1964).


\bibitem{crawford1964}
R. S. Harp and F. W. Crawford, J. Appl. Phys. {\bf 35}, 3436 (1964).


\bibitem{dote1965}
T. Dote and T. Ichimiya, J. Appl. Phys. {\bf 36}, 1866 (1965).


\bibitem{lepechinsky1966} 
D. Lepechinsky, A.M. Messiaen, and P. Polland, J. Nucl. Energy, Part C Plasma Phys. {\bf 8}, 165 (1966).


\bibitem{buckley1966}
R. Buckley, Proc. Roy. Soc. {\bf 290}, 186 (1966). 

\bibitem{balmain1966}
K. G. Balmain, Radio Sci. {\bf 1}, 1 (1966).

\bibitem{davis1966}
P.G. Davis, Proc. Roy. Soc. {\bf 88}, 1019 (1966). 

\bibitem{messiaen1966}
A.M.~Messiaen, P.E.~Vandenplas, J. Appl. Phys. {\bf 37}, 1718 (1966).


\bibitem{mckeown1967} D.L. McKeown and R.L. Ferrari, Int. J. Electron. {\bf 23}, 39 (1967). 


\bibitem{waletzko1967}
J. A. Waletzko and G. Bekefi, Radio Sci. {\bf 2}, 489 (1967). 


\bibitem{kostelnicek1968} R. J. Kostelnicek, Radio Sci. {\bf 3}, 319 (1968).

\bibitem{hellberg1968} M.A. Hellberg, J. Plasma. Phys. {\bf 2}, 395 (1968).

\bibitem{li1970} N.C. Li and W. A. Gustafson, Phys. Fluids {\bf 13}, 652 (1970). 


\bibitem{cohen1971} A. J. Cohen and G. Bekefi, Phys. Fluids {\bf 14}, 1512 (1971).
 

\bibitem{tarstrup1972}
J. Tarstrup and W.J. Heikkila, Radio Sci. {\bf 7}, 493 (1972).



\bibitem{aso1973}
T. Aso, Radio Sci. {\bf 8}, 139 (1973).

\bibitem{aso1973b} T. Aso, J. Geomag. Geoelectr. {\bf 25}, 325 (1973).


\bibitem{bantin1974}
C. C. Bantin and K. G. Balmain, Can. J. Phys. {\bf 52}, 291 (1974).


\bibitem{meyer1975} P. Meyer and N. Vernet, Radio Sci. {\bf 10}, 529 (1975).


\bibitem{vernet1975} N. Vernet, R. Manning, and J. L. Steinberg, Radio Sci. {\bf 10}, 517 (1975). 

\bibitem{stenzel1976} R.L. Stenzel, Rev. Sci. Instrum. {\bf 47}, 603 (1976).


\bibitem{nakatani1976} D.T. Nakatani and H. H. Kuehl, Radio Sci. {\bf 11}, 433 (1976).


\bibitem{kist1977}
R. Kist, Radio Sci. {\bf 12}, 921 (1977).


\bibitem{morin1991}
G.A. Morin and K.G. Balmain, Radio Sci. {\bf 26}, 459 (1991).


\bibitem{booth2005} S. Dine, J.P. Booth, G.A Curley, C.S. Corr, J. Jolly, and J. Guillon, Plasma Scources Sci. Technol. {\bf 14}, 777 (2005).

\bibitem{scharwitz2009}  C. Scharwitz, M. B\"oke, J. Winter, M. Lapke, T. Mussenbrock, and R. P. Brinkmann, Appl. Phys. Lett. {\bf 94}, 011502 (2009). 


\bibitem{xu2009} J. Xu, K. Nakamura, Q. Zhang, and H. Sugai, Plasma Sources Sci. Technol. {\bf 18}, 045009 (2009). 


\bibitem{xu2010} J. Xu, J. Shi, J. Zhang, Q. Zhang, K. Nakamura, and H. Sugai, Chinese Phys. B {\bf 19}, 075206 (2010). 


\bibitem{li2010} B. Li, H. Li, Z. Chen, J. Xie G. Feng, and W. Liu, Plasma Sci. Technol. {\bf 12}, 513 (2010).


\bibitem{liang2011} I. Linag, K. Nakamura, and H. Sugai, Appl. Phys. Express {\bf 4}, 066101 (2011)


\bibitem{schulz2015}
C. Schulz, T. Styrnoll, P. Awakowicz and I. Rolfes, IEEE Transactions on Instrumentation and Measurement {\bf 64}, 14981187 (2015)



\bibitem{blackwell2005}
D. D. Blackwell, D. N. Walker, and W. E. Amatucci, Rev. Sci. Instrum {\bf 76}, 023503 (2005).


\bibitem{hopkins2014}
M.A. Hopkins and L.B. King, Phys. Plasmas {\bf 21}, 053501 (2014).


\bibitem{sugai1999} H. Kokura, K. Nakamura, I.P. Ghanashev, and H. Sugai, Japan. J. Appl. Phys {\bf 38}, 5262 (1999).






\bibitem{lapke2008}  M. Lapke, T. Mussenbrock, and R. P. Brinkmann, Appl. Phys. Lett. { \bf 93}, 051502 (2008).





\bibitem{lapke2011} M. Lapke et al, Plasma Sources Sci. Technol. {\bf 20}, 042001, (2011). 





















\bibitem{lapke2013}
M. Lapke, J. Oberrath, T. Mussenbrock und R. P. Brinkmann, Plasma Scources Sci. Technol. {\bf 22}, 025005 (2013).

\bibitem{oberrath2014} 
J. Oberrath and R.P. Brinkmann, Plasma Sources Sci. Technol. {\bf 23}, 045006 (2014).


\bibitem{oberrath2014b} 
J. Oberrath and R.P. Brinkmann, Plasma Sources Sci. Technol. {\bf 23}, 065025 (2014).

\bibitem{friedrichs2018}
M. Friedrichs, J. Oberrath, EPJ Tech Instrum: Thematic Series on Novel Plasma Diagnostics {\bf 5}, 7 (2018).

\bibitem{oberrath2016} 
J. Oberrath and R.P. Brinkmann, Plasma Sources Sci. Technol. {\bf 25}, 065020 (2016).

\bibitem{oberrath2018} 
J. Oberrath, Plasma Sources Sci. Technol. {\bf 27}, 045003 (2018).


\bibitem{buckley1967}
R. Buckley, J. Plasma Phys. {\bf 1}, 171 (1967).

\bibitem{bernstein1959} 
I.B. Bernstein and I.N. Rabinowitz, Phys. Fluid {\bf 2}, 112 (1959).



















\end{thebibliography}
\end{document}